*Genome analysis*

# ggpicrust2: an R package for PICRUSt2 predicted functional profile analysis and visualization

Chen Yang[1], Jiahao Mai[1], Xuan Cao[2], Aaron Burberry[3], Fabio Cominelli[3,4], Liangliang Zhang[5, *]

[1]Department of Biostatistics, Southern Medical University, Guangzhou 510515, China

[2]Department of Mathematical Sciences, University of Cincinnati, Cincinnati, 45221, USA

[3]Department of Pathology, School of Medicine, Case Western Reserve University, Cleveland 44106, USA

[4]Case Digestive Health Research Institute, Case Western Reserve University, Cleveland 44016, USA

[5]Department of Population and Quantitative Health Sciences, Case Western Reserve University, Cleveland 44106, USA

*To whom correspondence should be addressed.

**Abstract**
**Summary:** Microbiome research is now moving beyond the compositional analysis of microbial taxa in a sample. Increasing evidence from large human microbiome studies suggests that functional consequences of changes in the intestinal microbiome may provide more power for studying their impact on inflammation and immune responses. Although 16S rRNA analysis is one of the most popular and a cost-effective method to profile the microbial compositions, marker-gene sequencing cannot provide direct information about the functional genes that are present in the genomes of community members. Bioinformatic tools have been developed to predict microbiome function with 16S rRNA gene data. Among them, PICRUSt2 has become one of the most popular functional profile prediction tools, which generates community-wide pathway abundances. However, no state-of-art inference tools are available to test the differences in pathway abundances between comparison groups. We have developed ggpicrust2, an R package, to do extensive differential abundance (DA) analyses and provide publishable visualization to highlight the signals.
**Availability and implementation:** The package is open-source under the MIT and file license and is available at CRAN and https://github.com/cafferychen777/ggpicrust2. Its shiny web is available at https://urlzs.com/EvDW8.
**Contact:** lxz716@case.edu
**Supplementary information:** Supplementary data are available at *Bioinformatics* online.

## 1 Introduction

One limitation of microbial community marker-gene sequencing is that it does not provide information about the functional composition of sample communities (Douglas *et al.*, 2020). By 2022, there are several methods available for predicting functions from the 16S rRNA sequence based on different approaches, such as PICRUSt2 (Douglas *et al.*, 2020), Tax4Fun2 (Wemheuer *et al.*, 2020), MicFunPred (Mongad *et al.*, 2021) and PICRUSt (Langille *et al.*, 2013) and so on. However, the accuracy and applicability of these methods depend on the specific research question and the characteristics of the microbial community being studied. Overall, these methods have greatly enhanced our ability to understand the functional roles of microbial communities in various environments, from the human gut to soil and water ecosystems. Among the various tools available, PICRUSt2 (Phylogenetic Investigation of Communities by Reconstruction of Unobserved States) has emerged as a highly favored instrument for predicting functional profiles, as it facilitates the generation of comprehensive pathway abundances within microbial communities. By doing so, PICRUSt2 provides researchers with valuable insights into the functional roles of microbial communities.

Nonetheless, a consensus regarding the optimal methodology for inferring and visualizing the functional abundance output generated by PICRUSt2 remains to be established within the academic community. As determining the statistically significant differences in functions and pathways between groups using Differential Abundance (DA) methods constitutes a critical step in the analysis, selecting an appropriate DA approach is indeed a topic of considerable importance within the scholarly discourse. The official wiki of PICRUSt2 initially recommended STAMP (Parks *et al.*, 2014) as the preferred software for analysis and visualization. However, STAMP has not been updated since 2015, indicating that it is unable to integrate the most recent advances in differential abundance (DA) analysis, which are crucial for systematically making statistical inferences from PICRUSt2 output data. Furthermore, STAMP presents installation challenges on macOS platforms, making it less user-friendly and potentially hindering its adoption by researchers in the field. The performance of five DA methods supported by STAMP, including ANOVA, Kruskal-Wallis H-test, t-test (equal variance), Welch's t-test, and White's non-parametric test, has been shown to be relatively inferior in a recent comparison of 20 DA methods across 38 datasets (Nearing *et al.*, 2022). The comparison concluded that AlDEx2 and ANCOM-II produce the most consistent results across studies and agree best with the interest of results from different approaches, but still recommend that researchers should use a consensus approach based on multiple differential abundance methods to help ensure robust biological interpretations (Nearing *et al.*, 2022). Despite there are several platforms or packages support the multiple advanced DA methods such as MicrobiomeAnalyst (Chong *et al.*, 2020), MicrobiomeExplorer (Reeder *et al.*, 2021), microbiomeMarker (Cao *et al.*, 2022), they are not specifically designed for PICRUSt functional output data. Due to the discrepancies in format and characteristics between PICRUSt2 output data and 16S rRNA genes data, the above platforms or software intended for the analysis of 16S rRNA genes data often encounter difficulties when importing PICRUSt2 data. Although almost all DA methods can be used in R, each method creates various burdens for data import and parameter configuration which increases both the effort and time cost, and diminishes the efficiency. Additionally, these R packages often lack the ability to visualize DA results and generate publication-quality figures. Thus, developing a user-friendly R package for analyzing PICRUSt2 functional output data using various DA methods and visualizations is urgently needed to fill the gaps.

## 2 *ggpicrust2* R package

The general workflow of the package is shown in Figure 1. *ggpicrust2* not only allows for recently developed advanced DA methods and visualization of results but also can convert PICRUSt2 output KO abundance tables into KEGG pathway abundance tables, which cannot be performed using PICRUSt2 alone. It also provides annotation of KO, EC, MetaCyc pathway, and KEGG pathway and enables classification of KEGG pathways. In the future, ggpicrust2 plans to incorporate a broader array of functional prediction tools, including but not limited to Tax4Fun2, in order to expand its capabilities and utility. Additionally, the package will integrate other methods that have demonstrated strong performance in simulation comparisons, ensuring continuous improvement and alignment with the latest advancements in the field.

### 2.1 Data input

*ggpicrust2* recommends adopting the data format of PICRUSt2 original output pred_metagenome_unstrat.tsv without reformat. But csv and txt is also acceptable. Furthermore, it is capable of supporting files that have been converted to mimic the PICRUSt2 output format, ensuring compatibility and flexibility for various data sources.

### 2.2 Conversion to KEGG pathway abundance

KEGG Orthology (KO) is a classification system developed by the Kyoto Encyclopedia of Genes and Genomes (KEGG) database (Kanehisa *et al.*, 2022). It uses a hierarchical structure to classify enzymes based on the reactions they catalyze. To better understand pathways' role in different groups and classify the pathways, the KO abundance table can be converted to KEGG pathway abundance. But PICRUSt2 removes the function from PICRUSt. *ko2kegg_abundance()* can help convert the table.

### 2.3 Advanced DA methods

Differential abundance (DA) analysis plays a major role in PICRUSt2 downstream analysis. *pathway_daa()* integrates almost all DA methods applicable to the predicted functional profile which there excludes ANCOM and ANCOMBC. It includes ALDEx2 (Fernandes *et al.*, 2013), DEseq2 (Love *et al.*, 2014), Maaslin2 (Mallick *et al.*, 2021), LinDA (Zhou *et al.*, 2022), edgeR (Robinson *et al.*, 2010), limma voom (Ritchie *et al.*, 2015), metagenomeSeq (Paulson *et al.*, 2013), lefser (Segata *et al.*, 2011) which have demonstrated varying degrees of success in distinct benchmarking assessments(Yang and Chen, 2022; Calgaro *et al.*, 2020; Nearing *et al.*, 2022).

### 2.4 Annotation of KO, EC, and pathway

*pathway_annotation()* can annotate the KO, EC, MetaCyc pathways' description from annotation table. And it can pull requests to online KEGG database to annotate KEGG pathways' pathway_name, pathway_description, pathway_class and pathway_map. The function can be used to annotate the output file of PICRUSt2 or the output table of pathway_daa().

### 2.5 Visualization

The mainstream visualization of PICRUSt2 is bar_plot, error_bar_plot, pca_plot, heatmap_plot. pathway_errorbar can show the relative abundance difference between groups and log2 fold change and p-values derived from DA results. *pathway_pca()* can show the difference after dimensional reduction via Principal Component Analysis (PCA). *pathway_heatmap()* can visualize the patterns in PICRUSt2 output data which can be useful for identifying trends or highlighting areas of interests.

### 2.6 Integration

*ggpicrust()* is the integration function of *pathway_daa()*, *pathway_annotation()*, *pathway_errorbar(), ko2kegg_abundance()*. This tool is designed to facilitate the entire data analysis process for those who are new to the field. However, it is also capable of being used by professional analysts in a modular fashion, allowing for increased customization and control. To further support users and promote the understanding of the package's capabilities, we have developed a detailed user manual, which is provided as Supplementary Materials. This document includes step-by-step installation instructions, explanations of the main features, and guidance on how to effectively leverage the *ggpicrust2* package within academic research. Our aim is to ensure that both novice and experienced researchers can easily access and benefit from the package's advanced functionalities.

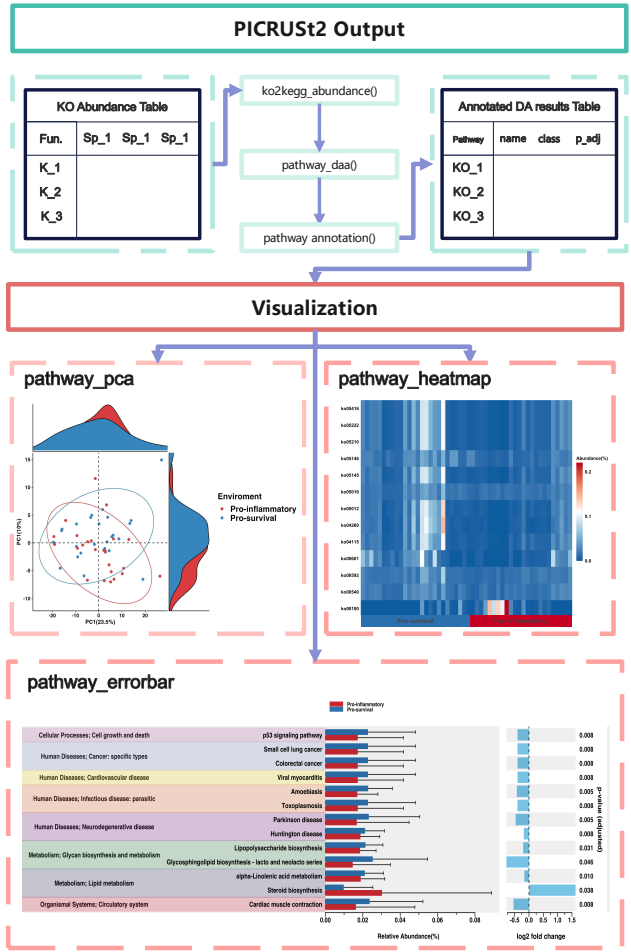

**Fig. 1.** Workflow and visualization example for the ggpicrust2 R package. The results of pathway_daa(), pathway_pca(), and pathway_heatmap() functions are depicted. The pathway_daa results reveal statistically significant differences in KEGG pathways between the two groups, which are classified and annotated according to pathway_class. The pathway_pca results display the Principal Component Analysis (PCA) findings, along with density plots for both groups. Finally, the pathway_heatmap results exhibit heatmaps for each group separately, showcasing the distinct patterns within the data.

### 2.7 Application

After employing PICRUSt2 to perform functional profile prediction on microbiome data from C9orf72 loss of function mice, including those that underwent fecal transplantation and those that did not (Burberry *et al.*, 2020), our subsequent data analysis using ggpicrust2 entailed the implementation of LinDA. This approach led to the identification of KEGG pathways that demonstrated statistically significant differences between the pro-survival and pro-inflammatory environments across both groups of mice. Of particular interest were the pathways ko05016, which is primarily involved in the pathogenesis of Huntington's disease, and ko05012, known for its association with Parkinson's disease. Both pathways are linked to human diseases and neurodegenerative disorders. The DA results were meticulously annotated, and the output was visualized for subsequent analysis. The visual representation of the results, which provides insights into the involvement of these pathways in the studied conditions, is depicted in Figure 1.

## 3 Conclusion

*ggpicrust2*, available at CRAN and https://github.com/cafferychen777/ggpicrust2, is an R package developed explicitly for PICRUSt2 predicted functional profile to do advanced differential abundance (DA) analysis and visualization of the DA results. This package effectively addresses the limitations of existing tools in terms of methods and visualization, and its integrated and distributed design caters to both professionals and beginners by meeting the needs of both groups. By providing a seamless experience for analyzing and visualizing DA results, *ggpicrust2* has the potential to significantly enhance the quality and efficiency of research involving functional profile predictions. *ggpicrust2* has already been incorporated into the PICRUSt2 wiki documentation, reflecting its growing recognition and adoption within the research community.

### Acknowledgments

We want to acknowledge Sonja Schaufelberger at University of Gothenburg for the feedback and suggestions regarding the *ggpicrust2* package. Her insights have significantly contributed to the improvement and development of our tool, ensuring that it remains both versatile and useful for researchers in the scientific community.

### Funding

This work has been supported by NIH grants 5P30AG072959-02 and 3R01DK042191-30S1.

*Conflict of Interest:* none declared.

*C.Yang et al.*